\documentclass[sigconf]{acmart}
\usepackage{booktabs} 
\usepackage{multirow}

\newcommand{\remove}[1]{}
\usepackage{xcolor} 
\usepackage{graphicx}
  \usepackage{subcaption}

\usepackage{dblfloatfix} 

\usepackage{comment}

\renewcommand\footnotetextcopyrightpermission[1]{} 
\setcopyright{none}
\acmDOI{}

\acmISBN{}
\acmPrice{}

\newcommand{\ts}{Minos}
\newcommand{\pvs}{\vspace{-8pt}}
\usepackage{algpseudocode}
\usepackage{algorithm}
\usepackage{enumitem}

\renewcommand\footnotetextcopyrightpermission[1]{} 
\begin{document}
\title{Size-aware Sharding For Improving Tail Latencies in In-memory Key-value Stores}

\author{Diego Didona, Willy Zwaenepoel}
\affiliation{
  \institution{EPFL}
  }

\begin{abstract}
This paper introduces the concept of size-aware sharding to improve tail latencies for in-memory key-value stores, and describes its implementation in the \ts{} key-value store. Tail latencies are crucial in distributed applications with high fan-out ratios, because overall response time is determined by the slowest response.

Size-aware sharding distributes requests for keys to cores according to the size of the item associated with the key. In particular, requests for small and large items are sent to disjoint subsets of cores. Size-aware sharding improves tail latencies by avoiding head-of-line blocking, in which a request for a small item gets queued behind a request for a large item. Alternative size-unaware approaches to sharding, such as keyhash-based sharding, request dispatching and stealing do not avoid head-of-line blocking, and therefore exhibit worse tail latencies.

The challenge in implementing size-aware sharding is to maintain high throughput by avoiding the cost of software dispatching and by achieving load balancing between different cores. \ts{} uses hardware dispatch for all requests for small items, which form the very large majority of all requests. It achieves load balancing by adapting the number of cores handling requests for small and large items to their relative presence in the workload.

We compare \ts{} to three state-of-the-art designs of in-memory KV stores. 
Compared to its closest competitor, Minos achieves a 99th percentile latency that is up to two orders of magnitude lower.  
 Put differently, for a given value for the 99th percentile latency equal to 10 times the mean service time, \ts{} achieves a throughput that is up to 7.4 times higher.

\remove{
This is not very good. It all of a sudden without warning starts talking about throughput, while all the previous discussion is about tail latency. Does not follow with the rest of the story. 

Need to fill in the numbers though. And copy the same sentences to the end of the introduction.

We demonstrate that \ts{} is able to meet strict tail latency requirements for throughput up to 7.4 higher than the second best  approach, and that \ts{} is able to autonomously adapt to changing workload conditions.
}

\end{abstract}

\settopmatter{printfolios=true}
\maketitle

\section{Introduction}
\label{sec:intro}
Many distributed applications use in-memory key-value (KV) stores as caches or as (non-persistent) data repositories~\cite{Nishtala:2013,Atikoglu:2012,Bronson:2013,memcached,Li:2017,Jin:2017}. 
Their performance, both in terms of throughput and latency, 
 is often critical to overall system performance.
Many of these applications exhibit a high fan-out pattern, i.e., they issues a large number of requests in parallel~\cite{Nishtala:2013}.
From the application's standpoint, the overall response time is then determined by the slowest of the responses to these requests, hence the crucial importance of tail latency for KV stores~\cite{Dean:2013}.

Given their importance, the performance of KV stores has been the subject of much recent work, both in terms of software and hardware.
Software optimizations include, among others, zero-copy user-level networking stacks, polling, run-to-completion processing, and sharding of requests between cores~\cite{Lim:2014,Ousterhout:2015,Kapoor:2012}.
Hardware optimizations primarily revolve around the use of RDMA~\cite{Kalia:2014,Kalia:2016}, programmable NICs~\cite{Kaufmann:2016,Li:2017} or GPUs~\cite{Zhang:2015,Hetherington:2015}.
The work reported in this paper does not require any particular hardware support. 
Instead, we assume commodity NICs with multiple queues and some mechanism to direct requests to a particular queue.

\pvs
~\\\noindent{\bf Variable item sizes and tail latency.} 
The workload observed for many KV stores consists of a very large number of requests for small items and a much smaller number of requests for large items~\cite{Atikoglu:2012,Nishtala:2013,Blott:2015}.
Because of their higher service times, however, handling the requests for larger items consumes a significant share of the available resources.
Processing these large items therefore increases the probability of head-of-line blocking, a situation in which a request for a small item ends up waiting while a large item is being processed.
As a result of the wait, that request experiences additional latency, which in turn may increase the tail latency of the KV store.
Even a very small number of requests for large items can significantly drive up tail latencies. 
More specifically, a percentage of large requests much smaller than N percent can lead to a considerable increase of the (100-N)th percentile. 

\pvs
~\\\noindent{\bf Size-aware sharding.} 
This paper introduces the notion of size-aware sharding to address this issue.
In general, size-aware sharding means that requests for items of different sizes go to different cores.  
In its simplest form,  it means that, for some cutoff value between small and large, small and large items are served by disjoint sets of cores.
The intuition behind size-aware sharding is that by isolating the requests for small items, they do not experience any head-of-line blocking, and, given that they account for a very large percentage of requests, the corresponding percentile of the latency distribution is improved.

The implementation of size-aware sharding poses several challenges.
A first challenge is how to continue to use hardware dispatch of an incoming request to the right core. In general, a client of the KV store does not know the size of an item to be read, and moreover it does not know which cores are responsible for small or large items. Therefore, size-aware sharding would seem to necessitate a software handoff in which an I/O core reads incoming requests and dispatches them to the proper core.
Instead, we demonstrate a method by which software dispatch is required only for the very small number of requests for large items.
Second, cutoff values between large and small items must be chosen and the proper number of cores must be allocated for handling small and large items. We show that, even in the presence of a workload that varies over time, this can be done by a simple control loop.

\pvs
~\\\noindent{\bf \ts{}.} This paper describes the \ts{} in-memory KV store that implements size-aware sharding using the above techniques. We compare \ts{} to alternative size-unaware designs based on keyhash-based request sharding, software handoff and work stealing, implemented by state-of-the-art systems such as MICA~\cite{Lim:2014}, RAMCloud~\cite{Ousterhout:2015} and ZygOS~\cite{Prekas:2017}. 

We show that \ts{} achieves a 99th percentile latency that is up to two orders of magnitude   lower than the second best approach. Put differently, for a given value for the 99th percentile latency equal to 10 times the mean service time, \ts{} achieves a throughput that is up to 7.4 times higher.

\pvs
~\\\noindent{\bf Contributions.} The contributions of this paper are:
\begin{itemize}[leftmargin=*]
\item the introduction of the notion of size-aware sharding for in-memory KV stores,
\item the design and implementation of the \ts{} KV store that implements size-aware sharding efficiently, and
\item the evaluation of \ts{} against state-of-the-art size-unaware designs.
\end{itemize}

\pvs
~\\\noindent{\bf Roadmap.} The outline of the rest of this paper is as follows. Section~\ref{sec:background} provides background on KV store workloads and discusses the shortcomings of existing approaches in achieving low tail latency. Section~\ref{sec:design} presents \ts{}' size-aware sharding approach. Section~\ref{sec:impl} discusses implementation details. Section~\ref{sec:testbed} describes the experimental environment. Section~\ref{sec:eval} presents experimental results. Section~\ref{sec:rw} discusses related work. Section~\ref{sec:conclusion} concludes the paper.
\section{Background}
\label{sec:background}
\begin{figure}[t!]
       \includegraphics[scale = 0.5]{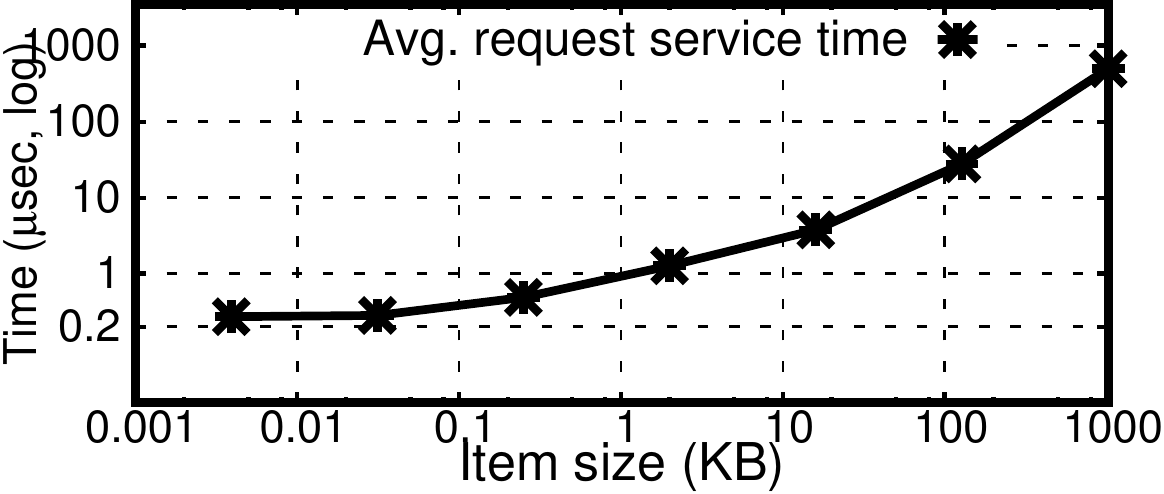}
        \caption{Service time of GET operations on items of different sizes on our platform (y axis in log scale). The service time measures the interval from the reception of the client request on the server to the transmission of the reply message. To avoid queueing effects, only a single client performs operations in a closed loop. The time to process a large item can be up to almost four orders of magnitude higher than what is needed to serve a small one.}
        \label{fig:background:service}
\end{figure}
\subsection{Item Sizes in Production KV Workloads}
\label{sec:background:production}
The sizes of the items stored and manipulated by KV stores in production environments can span orders of magnitude. For instance, large variations in item size have been reported in several deployments of the popular \texttt{memcached} KV store~\cite{memcached}. The Facebook \texttt{ETC} \texttt{memcached} pool stores items that vary in size from a handful of bytes to 1 Mbyte~\cite{Atikoglu:2012}. The size distribution is heavy-tailed: the 5th percentile in the \texttt{regional} pool is 231 bytes, while the 99th percentile is 381KB~\cite{Nishtala:2013}. A similar degree of variability in item size has also been reported for other KV deployments such as Wikipedia~\cite{Lim:2013} and Flickr~\cite{Blott:2015}, where item sizes span up to 4 orders of magnitude, from 500B to 1 MB.

Moreover, Atikoglu et al. report that in the \texttt{ETC} \texttt{memcached} pool at Facebook requests for large items, despite being rare, consume a large share of  the computational resources, because service times are closely related to item size, and account for a significant fraction    of the transfered data~\cite{Atikoglu:2012}. 
   This dynamic is consistent with observations from similar application domains, such as, e.g., web servers~\cite{Arlitt:1997,Crovella:1997} and large-scale clusters~\cite{Wang04}.

\subsection{Variations in Item Size and Tail Latencies}
\label{sec:background:hob}

\begin{figure*}[t!]
\begin{subfigure}[b]{0.3\textwidth}  
       \includegraphics[scale = 0.5]{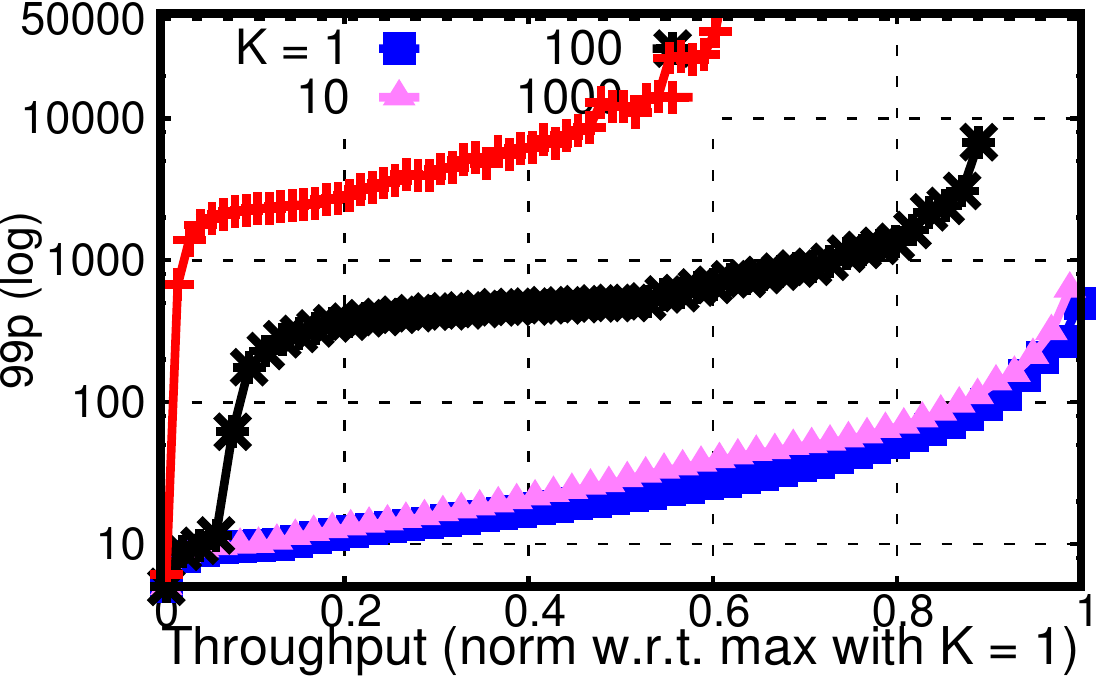}\caption{nxM/G/1.}\label{fig:background:99p:mica}
\end{subfigure}
\hfill
\begin{subfigure}[b]{0.3\textwidth}
       \includegraphics[scale = 0.5]{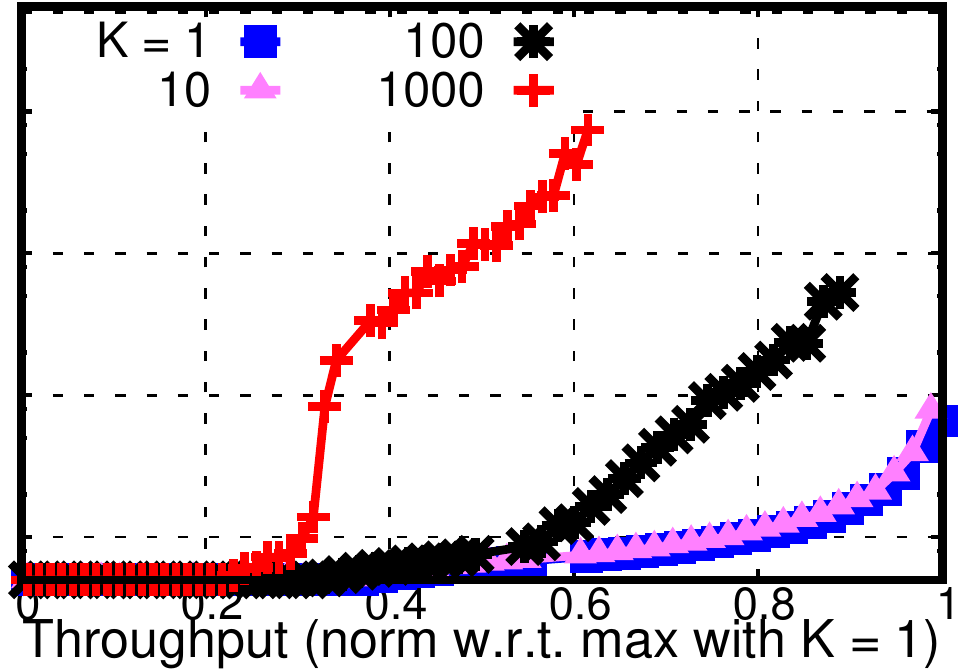}\caption{M/G/n.}\label{fig:background:99p:ramcloud}
\end{subfigure}
\hfill
\begin{subfigure}[b]{0.3\textwidth}
       \includegraphics[scale = 0.5]{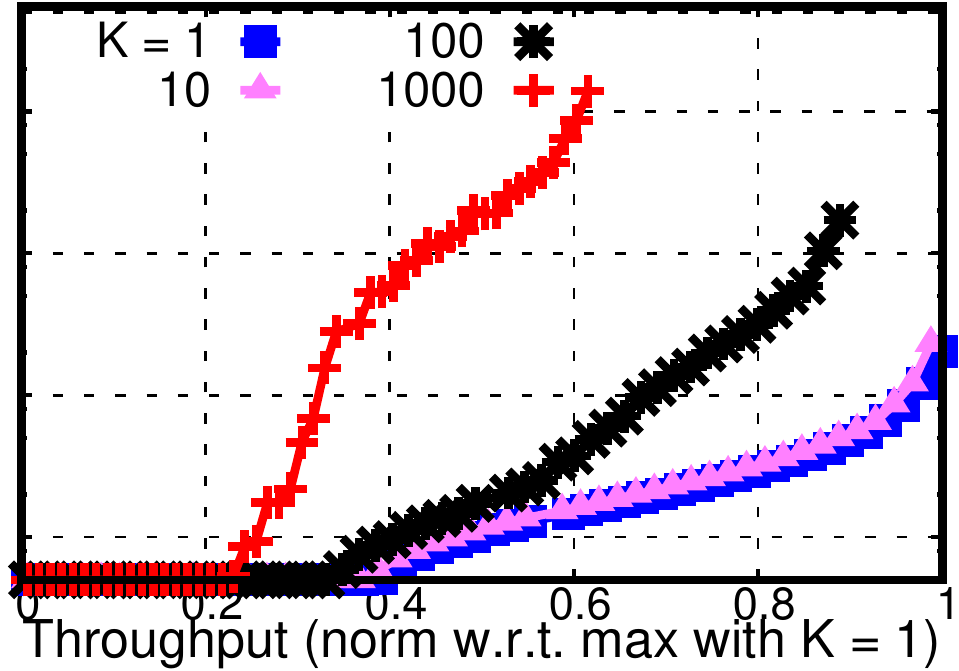}\caption{nxM/G/1 + work stealing.}\label{fig:background:99p:zygos}
\end{subfigure}
        \caption{Throughput vs. 99th percentile of response times for different type of queues (y axis in log scale). The service time distribution is bimodal: 0.125\% of requests is for large items; the remaining is for small ones. A large request has a service time K times larger than a small one. K is varied from 1 to 1,000. A small (<1\%) fraction of large requests suffices to hamper greatly the 99th percentile of response times, and to considerably reduce the achievable throughput.}
        \label{fig:background:99p}
\end{figure*}

Variations in item size have profound implications for tail latencies. As anecdotal evidence, Nishtala et al. report that in the Facebook \texttt{memcached} servers the median response time is 333 microseconds, while the 95th percentile is 1.135 milliseconds~\cite{Nishtala:2013}. In this section we show that this finding goes beyond the anecdotal, and that all common size-unaware sharding techniques exhibit high tail latencies for workloads in which even only a small fraction of requests targets large items. In particular, we show that, even under moderate loads, the (100-N)-th percentile is affected dramatically by a fraction, much smaller than N\%, of requests for large items. In the following we report on the 99th percentile, commonly used in Service Level Objective (SLO) definitions, but the results apply  
 to other high percentiles as well.

We simulate the operation of three common size-unaware sharding techniques on a server with n cores: 
\begin{itemize}[leftmargin=*]
\item {\bf Multiple queues (nxM/G/1)}, 
 where requests are dispatched immediately (early binding) to a queue for a particular core, often based on a keyhash, similar to what is used, for instance, in the EREW version of MICA~\cite{Lim:2014}.
\item {\bf Single-queue (M/G/n)}, in which requests are kept in a single queue and dispatched to a core when it becomes idle (late binding), similar to what is used, for instance, in RAMCloud~\cite{Ousterhout:2015}.
\item {\bf Multiple queues augmented with work stealing}, where requests are handled as in nxM/G/1, but in addition idle cores steal requests from the queues of other cores, similar to what is used, for instance, in ZygOS~\cite{Prekas:2017}. 
\end{itemize}

For simplicity, we use a workload with a bimodal size distribution. 
Small requests form 99.875\% of the workload, and have a service time of 1 time unit. Large requests form the remaining 0.125\%. We run different simulations in which the service time of these large requests is, respectively, K = 10, 100 and 1,000 time units. These values are in line with the order-of-magnitude differences in service time between small and large items observed on our platform (See Figure~\ref{fig:background:service} for a graph that depicts service time as a function of item size). 
 Inter-arrival times follow an exponential distribution. 
We furthermore assume an idealized scenario with zero overhead for dispatching requests to cores, no need for synchronization, and no adverse effects from lack of locality. 

We stress that our goal with this simulation is {\em not} to predict differences between these strategies in any real implementation, as their  
 performance in practice is greatly affected by various considerations such as locality, cost of synchronization, and cost of dispatching, which are not modeled in this simulation. 
Instead, our goal is to demonstrate, for all three methods, the substantial increase in tail latency as a result of the presence of a small fraction of requests for large items. 

Figure~\ref{fig:background:99p} shows the 99th percentiles for the three sharding strategies under this workload compared to a workload with an identical offered load, but with only requests for small items. 
Even though the fraction of large items requested is much smaller than 1 percent, Figure~\ref{fig:background:99p} shows a very considerable increase in the 99th percentile latency for all three strategies. {\color{black}For K = 100 and K = 1,000, at only 10\% utilization,  the 99th percentile for nxM/G/1 is one to two orders of magnitude higher than the 99th percentile in the workload composed only of small requests. Stealing and late binding are more resilient to service time variability at low load. As the load grows, however, they also suffer from one or two orders of magnitude degradation of the 99th percentile, with respect to the workload with only small requests.}

While all strategies produce increases in the 99th percentile, the reasons for these increases are somewhat different from one strategy to the next.

The nxM/G/1 strategy suffers from head-of-line blocking when a request for a small item ends up in a queue behind a request for a large item or behind a request for a large item being executed by this core. 

The late binding of requests to cores makes M/G/n more resilient against head-of-line blocking than nxM/G/1, a well known result from queueing theory~\cite{Harchol-Balter:2013:book}, but it does not completely avoid it. The  nxM/G/1 strategy is vulnerable to cases in which the arrival of many large requests in a short period of time leads many (or even all) cores to be busy serving large requests. Such an event temporarily reduces the amount of resources available to serve small requests, which impacts tail latency.

Stealing improves the tail latency of nxM/G/1, as it steals some of the requests that would otherwise experience head-of-line blocking   
 but it cannot completely avoid head-of-line blocking.  First, stealing only occurs when a core is idle, and the likelihood of a core being idle decreases as the load increases. Second, by the time a core becomes idle, 
 a request that it steals is likely to have already experienced some head-of-line blocking in the queue from which it is stolen.

\pvs
~\\
In light of these results, \ts{} processes requests for small and large items on disjoint set of cores, a technique we call  {\em size-aware sharding}. This addresses the shortcomings of existing approaches, by avoiding that a request on a small item waits for the completion of a request on a large one. 

\section{Size-aware sharding in \ts{}}
\label{sec:design}
\noindent{\bf Preliminaries.} We consider a server with $n$ cores.
The server has a multi-queue NIC, with multiple receive (RX) and transmit (TX) queues.  

 We configure the NIC with $n$ RX queues and $n$ TX queues.
At any time, 
 there are $n_l$ cores handling requests for large items and $n_s$ cores handling requests for small items ($n_l$ + $n_s$ = $n$).
With a slight abuse of language, we say that a request for a small (large) item is a small (large) request, and that a core handling small (large) requests is a small (large) core. 
In addition to an RX and a TX queue, each large core maintains a software queue.

In the following, we assume all $n$ cores are within the same NUMA domain, so that KV item accesses and inter-core communication happen within the same NUMA domain. \ts{} can seamlessly scale to multiple NUMA domains by running an independent set of small and large cores within each NUMA domain, and by having clients send requests to the NUMA domain that stores the target key~\cite{Lim:2014}.

We consider a KV store with the usual CRUD (Create, Read, Update, Delete) semantics. 
A client can perform a GET(key) and a PUT(key, value).
Create and delete are considered special versions of PUT, and not discussed any further.
When a client issues GET and PUT operations, the client software puts in the request the id of the RX queue 
in which the corresponding packets are deposited when they arrive at the server. The target RX queue is chosen at random for GET operations, and depends on the keyhash for PUT operations (as we describe in Section~\ref{sec:impl:kvs}).
A PUT request also includes the size of the item that is being written.
The client does not know the size of an item to be read.
Furthermore, the client does not need to know which or how many cores on the server handle small or large requests.  

In the following discussion we initially assume that we know the threshold on the item size that separates small and large items. We explain later how the actual threshold is determined. We first explain size-aware sharding with a given number of small cores and one large core. We explain later how the number of small and large cores is determined, and how the system operates with a number of large cores different from one.

\pvs
~\\\noindent{\bf Receiving incoming requests.}
Only the small cores read incoming requests from the RX queues. 
They do so in batches, to amortize the cost of communicating with the NIC.
Each small core repeats the following sequence of actions w.r.t. the RX queues: First, it reads a batch of B requests from its own RX queue. 
Then it reads a batch of B/$n_s$ requests from the RX queue of the large core.
In this way, all RX queues are drained at approximately the same rate.
The reason a large core never reads incoming requests from its RX queue is that, if it were to receive a small request, this request could experience head-of-line blocking behind large requests.

We start by explaining how GET operations are handled.

\pvs
~\\\noindent{\bf Operation of the small cores.} For each request, a small core looks up the item associated with the requested key. If its size is below the threshold, the small core continues the GET operation and replies to the client with the requested item (by putting the corresponding reply packet(s) on its TX queue). 
 If the length is above the threshold, the small core puts the request in the software queue of the large core.

\pvs
~\\\noindent{\bf Operation of large core.} A large core looks at its software queue. If it finds an incoming request, it finds the corresponding item, and replies to the client by putting the item in its TX queue. 

\pvs
~\\
The operation of a PUT is mostly similar, except that the size is known to the client and present in the request. There is therefore no need to do a lookup to find the size, and, depending on the new size, the request is handled either immediately by a small core or passed on by a small core to the large core and handled there.

\pvs
~\\\noindent{\bf How to find the threshold between large and small.}  Each small core maintains a histogram of the number of requests that correspond to item sizes in certain ranges. This histogram is updated on the receipt of every request according to the size of the target item. Periodically, core 0 aggregates these histograms, finds the size corresponding to the 99th percentile, declares that size to be the threshold for the next epoch, and resets the histograms to zero.

To be resilient to transient workload oscillations, core 0 smooths the values in the aggregated histogram (noted $H$) according to a moving average that uses the histogram obtained in the previous epoch (noted $H_{curr}$). That is, for each entry $i$, core 0 computes $H_{curr}[i] = (1-\alpha) H_{curr}[i]  \cdot \alpha H[i] $, and uses the new $H_{curr}$ to determine the 99th percentile. $\alpha$ is a discount factor in the range [0,1], and determines the weight of the new measurements over previous ones. Because \ts{} targets high throughput workloads, many item sizes are sampled during an epoch. Hence, $H$ is highly representative of the current workload, and is assigned a weight equal to 0.9~\cite{Zhang:2005}.

\pvs
~\\\noindent{\bf How to choose the number of small cores.} We maintain a cost function that gives us for a request of a given size a certain processing cost. \ts{} can use various cost functions, but currently uses the number of network packets handled to serve the request as cost, either the number of packets in an incoming PUT request or the number of packets in an outgoing GET reply.  Alternatives would be the number of bytes or a constant plus the number of bytes. In any case, the fraction of cores that serve as small cores is set to the ceiling of the fraction of the total processing cost incurred by small requests times the total number of cores. The remaining cores are used as large cores.

\pvs
~\\\noindent{\bf Operating with a number of large cores different from one.} If, as a result of the above calculation, there is more than one large core, then \ts{} distributes the large requests over the large cores such that each large core handles a non-overlapping contiguous size range of requests, and such that the processing cost of requests assigned to each large core is the same.  By doing so, not only does Minos balance the load on large cores, but it also shards large requests in a size-aware fashion. That is, the smallest among the large requests are assigned to the first large core, and larger requests are progressively assigned to other cores. Each large core has a software queue, and a small core that receives a large request puts the request in the software queue of the large core that is handling the size of the requested item. 

If all cores are deemed to be small cores, then one core is designated a standby large core. In other words, it handles small requests, but if a large request arrives, it is sent to this core, which then becomes a large core.

\pvs
~\\\noindent{\bf Design rationale.} The goal of \ts{} is to improve the 99th percentile. To that end we identify the smallest 99 percent of the requests. We isolate the processing of these requests from the processing of larger requests, such that no head-of-line blocking occurs. Furthermore, we assign a sufficient number of cores to handle these requests such that no long request queues can materialize for these cores. 

The use of randomization and of the hashed value of the key to decide the target RX queue for a request leads to reasonable load balance among the RX queues. A similar observation was made in the context of MICA~\cite{Lim:2014}.
Since the small cores handle the requests that arrive in their own RX queue, and an equal portion of the requests that arrive in the RX queues of the large cores, overall the load is balanced among the small cores. 
By using purely hardware dispatch for the small requests we eliminate any unnecessary overhead in their processing, such as, for instance, software dispatches. 
We achieve these results while never dropping large requests, since there is always at least one core available for handling large requests. 

The only overheads compared to a purely hardware dispatch solution such as MICA are then:
1) software dispatch for the very small number of large requests,
2) synchronization on the RX queue and the software queue of the large cores, for which we found contention to be low, and
3) some minor loss in locality for the small requests that arrive in the RX queues of large cores.

\section{Implementation}
\label{sec:impl}
\subsection{Network stack}
\label{sec:impl:net}
\ts{} relies on the availability of a multi-queue NIC with support for redirecting, in hardware, a packet to a specific queue on the NIC (e.g., RSS~\cite{rss} or Flow Director~\cite{Fdir:intel}).
This feature is now commonplace in commodity NICs.

To reduce packet processing overhead, \ts{} uses the Intel DPDK library~\cite{dpdk} to implement a user-level zero-copy network stack. 
All memory for the DPDK library is statically allocated and accessible by all cores. 
Packets are received directly in memory, thus enabling  zero-copy packet processing. 
Furthermore, \ts{} uses DPDK-provided lockless software rings to dispatch large requests from small to large cores
without any copies~\cite{dpdk-rings}.
Small cores check for incoming requests by means of polling, to avoid costly interrupts~\cite{Ousterhout:2015}.
Similarly, large cores use polling to check for incoming requests on their software queue.
Requests are moved in batches to further limit overhead.

Communication between clients and servers uses UDP, implemented on top of Ethernet and IP.
Clients use the UDP header to specify the target RX queue for a given packet. 
Requests that span multiple frames (large PUT requests and large GET replies) 
are fragmented and defragmented at the UDP level. 

Retransmission is handled by the client. Similar to previous work~\cite{Lim:2014}, \ts{} does not support exactly-once semantics and assumes idempotent operations. Guaranteeing exactly-once semantics can be achieved by means of request identifiers.

\subsection{KV store and memory management}
\label{sec:impl:kvs}
\noindent{\bf Data structures.} \ts{} employs the KV data structures used in MICA~\cite{Lim:2014}. 
Keys are split in {\em partitions}.
Each partition is a hash table, each entry of which points to a bucket, equal in size to a cache line. 
Each bucket contains a number of slots, each of which contains a tag and a pointer to a key-value item. 
A first portion of the keyhash is used to determine the partition, a second portion to map a key to a bucket
within a partition, and a third portion forms the tag~\cite{Fan:2013}, which is used to reduce the number of random memory accesses when performing a key lookup~\cite{Lim:2014}.  
Overflow buckets are dynamically assigned to a bucket when it has reached its maximum capacity.

\pvs
~\\\noindent{\bf Memory management.} The current prototype of \ts{} employs the memory manager of the DPDK library to handle allocation of memory regions for key-value entries. \ts{} can be extended to integrate more efficient memory allocators, such as the one based on segregated fits of MICA, or a dynamic one as in Facebook's \texttt{memcached} deployment~\cite{Nishtala:2013}.

\pvs
~\\\noindent{\bf Concurrency control scheme.} \ts{} uses a concurrency control scheme that is similar to Concurrent Read Exclusive Write (CREW)~\cite{Lim:2014}. In CREW, each core is the {\em master} of one partition, and a given key can be written only by the master core of corresponding partition. This naturally serializes write operations on a key.  

The concurrency control scheme in \ts{} differs slightly from CREW, as a result of the distinction between small and large cores.
PUTs on keys whose master core is a small core proceed along the lines of CREW.
PUTs on keys whose master core is a large core may be served by any core (either because the request is small, or because it is dispatched to a large core different from the one which receives the request). Hence, PUTs are guarded by a spinlock.

We argue (and we experimentally show) that the corresponding overhead is largely outweighed by the benefits of size-aware 
sharding, especially for the read-dominated workloads that are prevalent in production environments~\cite{Atikoglu:2012,Nishtala:2013,Bronson:2013,Noghabi:2016}. 
First, in such workloads PUTs are rare.
Second, PUTs on large cores proceed without contention, because large cores serve non-overlapping size ranges, so requests for the same large item are sent to the same core.
Third, PUTs on small cores mostly proceed without contention because of the CREW nature of the concurrency protocol
for keys whose master is a small core.

GETs can be served by any core, and are
served by means of an optimistic scheme~\cite{Lim:2014}. 
Each bucket has a 64-bit epoch, which is incremented when starting and ending a write on a key stored in that bucket.
Upon reading, a core looks at the epoch. 
If it is odd, then there is an ongoing write on a key of the bucket, and the read is stalled until the epoch becomes even.
If (or when) the epoch is even, the core saves the current epoch value and performs the read. 
After the read, the core re-reads the epoch of the bucket. 
If the value is the same as when the read started, the read is successful. 
Else, a conflicting write might have taken place, and the read is restarted.

\section{Experimental Platform}
\label{sec:testbed}
\subsection{Hardware}
\label{sec:testbed:platform}
Our experimental platform is composed of 8 identical machines equipped with an Intel(R) Xeon(R) CPU E5-2630 v3 @ 2.40GHz with 8 physical cores and 64 GB of main memory.  The machines run Ubuntu 16.04.2 with a 4.4.0-72-generic kernel. One machine acts as server and the other 7 run the client processes. We disable hyperthreading and power-saving modes on all the machines. All the machines are equipped with a 40Gbit Mellanox MT27520 NIC (ConnectX-3 Pro), are located in the same physical rack, and are connected via a top-of-rack switch. The network stack for both client and server machines relies on the Intel DPDK library (version 17.02.1), to which we allocate 50 1GB huge pages. 

Our NIC supports only RSS to implement hardware packet-to-RX queue redirection~\cite{mlx4:rss}. RSS determines the RX queue for an incoming packet by performing the hash of the quintuplet composed of source and destination IP, source and destination port and the transport layer protocol. To allow the clients and the server to send packets to specific RX queues, we ran a set of preliminary experiments to determine to which port to send a packet so that it is received by a specific RX queue. More flexible hardware packet redirection methods can be used on NICs that support them. For example \ts{} can use Flow Director~\cite{Fdir:intel,Fdir:mellanox} to set the target RX queue as UDP destination port of a packet.

\subsection{Systems used in comparison}
We compare \ts{} with three systems that implement state-of-the-art designs of KV store, and that are based on the queueing models that we have described in  Section~\ref{sec:background}. 

\begin{itemize}[leftmargin=*]
\item {\bf Hardware Keyhash-based sharding (HKH).} This system implements the  nxM/G/1 queueing model, as done in MICA~\cite{Lim:2014}. Requests are redirected in hardware to the target core, according to the CREW policy.
This policy performs the best on skewed read-dominated workloads~\cite{Lim:2014}, such as our default workload. 
\item {\bf Software hand-off (SHO).} This system implements the  M/G/n queueing model, as in RAMCloud~\cite{Ousterhout:2015}. SHO uses disjoint sets of handoff and worker cores. Each handoff core has a software queue, in which it deposits the requests taken from its RX queue. Worker cores pull one request at a time from the handoff queues (in round robin if there is more than one), process the corresponding KV request, and reply to the client. The number of handoff cores is fixed and known {\em a priori} by the clients, which only send requests to the corresponding RX queues. The throughput of SHO is bounded by the dispatch rate of handoff cores. The best number of handoff cores depends on whether the workload is CPU or network bound. We have experimented with 1,2 and 3 handoff cores. We report experimental results corresponding to the best configuration for each workload.

\item {\bf HKH + work stealing (HKH+WS).}  This system implements request stealing on top of HKH, as in ZygOS~\cite{Prekas:2017}. Each core has a software queue in which it places the requests taken from its own RX queue. When a core is idle, it steals requests from the software queues of other cores. If or when all software queues are empty, an idle core steals requests from another RX core's queue. Between stealing attempts, a core checks whether it has received any new request. If it has, it stops stealing and processes its own requests.
Cores steal requests from the software queues of other cores one at the time. Batching could introduce head-of-line blocking if the batch contains a large request followed by a short request, and is therefore not used. However, packets are stolen from other RX queues in batches, to increase resource efficiency. Requests stolen from another core's RX queue are put in the stealing core's software queue, so they can be stolen in turn.
\end{itemize}

\remove{
We further implement a system that implements our size-aware sharding on top of SHO (SHO+SA).
Requests of different sizes are dispatched to worker cores according to the same load balancing technique used by \ts{} to distribute large requests (Section~\ref{}). To obtain the size of a GET operation,  a handoff core performs a lookup operation from the KV store. We use SHO+SA to show that our size-aware sharding technique can be beneficial also for systems that, by design, rely on software handoff (e.g., RamCloud~\cite{Ousterhout:2015} and Rocksteady~\cite{Kulkarni:2017}).
}

\remove{
SHO and SHO+SA can be configured to use a different number of handoff cores. In general, when the load posed by large requests is high, one handoff core is enough to keep up with the requests arrival rate, and all other cores can process KV operations. As the bulk of the load shifts to small requests, more handoff cores are needed to keep up with the higher arrival rate.
We have experimented with 1, 2 and 3 handoff cores. Using 3 handoff cores always resulted in worse performance because it results in using only 5 cores to serve KV operations. 
 In the next section, we report experimental results corresponding to the best configuration for each workload and target SLO.
 }

For a fair comparison, all the designs we consider are implemented in the same codebase. In particular, they all use the same KV data structure (~\S~\ref{sec:impl:kvs}) and lightweight network stack (~\S~\ref{sec:impl:net}). 

The internal parameters of \ts{} are set as follows. Workload statistics are collected by core 0 every second. The size of a batch of requests read from a RX queue is 32, and the same batch size is used for other systems as well.

\subsection{Workloads}
\label{sec:testbed:wkld}
\begin{table}[t]
\footnotesize
\centering
\begin{tabular}{|c|c|c|}
\cline{1-3}
{\bf \% large reqs ($p_L$)}       & {\bf Max size ($s_L$)}  & {\bf \% data for large reqs}     \\ \cline{1-3}
\multirow{3}{*}{0.125} & 250 KB                  & 25   \\ \cline{2-3} 
                         & 500 KB                  & 40   \\ \cline{2-3} 
                         & 1000 KB                 & 60   \\ \hline
0.0625                 & \multirow{4}{*}{500 KB} &  25    \\ \cline{1-1} \cline{3-3} 
0.25                   &                         & 60   \\ \cline{1-1} \cline{3-3} 
0.5                    &                         & 75   \\ \cline{1-1} \cline{3-3} 
0.75                   &                         & 80   \\ \hline
\end{tabular}
\caption{Item size variability profiles.}
\label{tab:wkld}
\end{table}
We use workloads characterized by different degrees of item size variability and GET:PUT ratios.

\pvs
~\\\noindent{\bf Item size variability.} We use, as a starting point, the characterization of the ETC workload at Facebook~\cite{Atikoglu:2012}. Specifically, we consider a trimodal item size distribution, according to which an item can be tiny (1-13 bytes), small (14-1400 bytes) or large (1500-maximum size). The size of a specific item within each class is drawn uniformly at random. To generate workloads with different degrees of item size variability, we vary both the percentage of large requests, (noted $p_L$), and the size of items corresponding to large requests, by changing the maximum size of large items (noted $s_L$). 
We let $s_L$ range from 250KB to 1MB. These values are consistent with the production workloads we discussed in Section~\ref{sec:background:production}.  Because we focus on 99th percentile response times, we set $p_L < 1\%$, so that the 99th percentile of the requests service times corresponds to small and tiny items only. Specifically, we vary $p_L$ from 0.0625 to 0.75. 

Table~\ref{tab:wkld}  reports the combinations of $p_L$ and $s_L$ we consider. It also reports the corresponding percentage of bytes that are exchanged because of large requests.

\pvs
~\\\noindent{\bf Key popularity.}
We consider a skewed workload that follows a zipfian distribution with parameter 0.99. This represents the default value in YCSB~\cite{Cooper:2010}, is widely used in the evaluation of several KV stores~\cite{Lim:2014,Kalia:2014}, and is representative of the strong skew of many production workloads~\cite{Atikoglu:2012,Balmau:2017}. 

We use the zipfian distribution on the sets of tiny and small items, because they are many and they exhibit small variability in size. Large items, instead, are much fewer and exhibit much higher variability, and are therefore chosen uniformly at random. This avoids pathological cases in which the most accessed large item is the biggest or the smallest item, 
thereby skewing the results. 

We consider a dataset of 16M key-value pairs, out of which 10K are large elements. Of the remaining key-value pairs, 40\% correspond to tiny items, and 60\% to small ones. This setting is consistent with the item size distribution and the low access probability of individual large keys that characterize the ETC workload. Each large item has, in fact, a probability $p_L/100 \cdot  10K/16M$ of being accessed.
For simplicity, we keep the size of the keys constant to 8 bytes.

\pvs
~\\\noindent{\bf Write intensity.} We consider a read-dominated and a write-intensive workloads, corresponding, respectively, to a 95:5 and 50:50 GET:PUT ratio. These values are used as default values in YCSB and KV store evaluations~\cite{Lim:2014,Kalia:2014}. Moreover, in the ETC workload, 97\% of requests are GET operations.

\pvs
~\\\noindent{\bf Default workload.} We define one default value for each experiment parameter. We generate additional workloads by changing the value of one parameter at a time while keeping the other ones to their default values.
The default workload we consider is skewed with a 95:5 GET:PUT ratio, a percentage  of large requests equal to 0.125\% and a maximum large item size of 500 KB.

\subsection{Benchmarking methodology} 
\noindent{\bf Load generation.} We spawn 8 threads per client machine, each pinned to a separate physical core and to an RX queue. Client threads simulate an open system by generating requests at a given rate, which varies depending on the target arrival rate. The time between two consecutive requests of a thread is exponentially distributed.

\pvs
~\\\noindent{\bf Measurements.} 
Each request is timestamped with the send time at the client, which is piggybacked by the server on the reply message. Client threads constantly check their own RX queues  for replies, and compute the end-to-end latency of a request using the 
timestamp in the reply message.

{\color{black}A client thread can have multiple requests in flight, so for simplicity packet retransmission is not enabled. For this reason, we only report performance values corresponding to scenarios in which the packet loss rate is equal to 0.

Each workload runs for 60 seconds. The first and last 10 seconds are not included in the reported results.

\pvs
~\\\noindent{\bf Performance metrics.} We focus on maximum achievable throughput and 99th percentile of end-to-end latencies. We also measure the utilization of the server NIC to evaluate whether \ts{} is able to fully use the available bandwidth.

We consider SLOs in the form ``The 99th percentile of latencies must be within X times the mean request service time". On our platform and for default workload the mean service time is 5 $\mu$sec. We set X to 10 and 20, i.e., the target 99th percentile latency values to 50 and 100 $\mu$sec.  X = 10 corresponds to a strict SLO, and is the same value used in the evaluation of Zygos~\cite{Prekas:2017}. X = 20 corresponds to a looser SLO, and we use it to evaluate the sensitivity of \ts{} performance gains as a function of the strictness of the SLO.

\begin{figure}[t!] 
\centering
       \includegraphics[scale = 0.55]{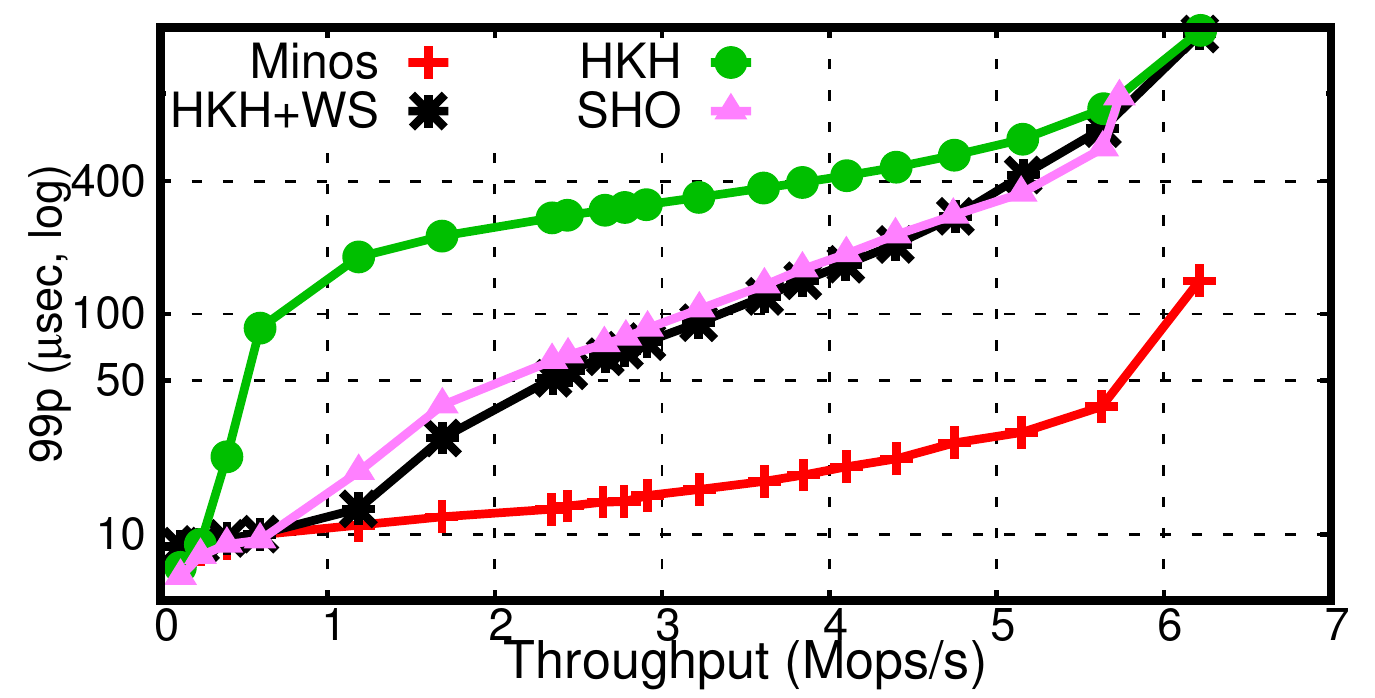}\caption{Throughput vs. 99th percentile latency (y axis in log scale) with the default workload. By efficiently separating small and large requests, \ts{} is able to deliver the highest throughput and the lowest tail latency. \ts{} matches the throughput of the purely hardware-based design and achieves tail latencies lower than the software handoff design.}\label{fig:eval:default}
\end{figure}

\section{Evaluation}
\label{sec:eval}

~\noindent{\bf Summary.} The highlights of our evaluation are as follows.
\begin{itemize}[leftmargin=*]
\item \ts{} achieves both low latency and high throughput. Compared to its closest competitor, \ts{} 
achieves a 99th percentile latency that is one to two orders of magnitude lower (\S~\ref{sec:eval:default}, \S~\ref{sec:eval:dynamic}).
With a 99th percentile specified to be 10 times the mean service time, its throughput is up to 7.4 times higher than the second best approach (\S~\ref{sec:eval:variability}).

\item \ts{} achieves good performance under both read-intensive and write-intensive workloads (\S~\ref{sec:eval:write}).

\item \ts{}  scales with the amount of available network bandwidth (\S~\ref{sec:eval:bandwidth}).

\item \ts{} achieves load balancing across cores (\S~\ref{sec:eval:lb}). 

\item \ts{} can adapt to changing workload conditions (\S~\ref{sec:eval:dynamic}).

\end{itemize}

\subsection{Default workload}
\label{sec:eval:default}
\noindent{\bf Throughput vs. 99th percentile latency.} Figure~\ref{fig:eval:default} shows the 99th percentile latency (99p) as a function of the throughput with the default workload. 
\ts{} achieves the best peak throughput (6.2 Mops) and the lowest latency ($\leq 50 \mu sec$ up to 90\% of peak throughput). 

\ts{} achieves similar peak throughput as HKH and HKH+WS, reflecting the fact that all three systems rely mostly or entirely on hardware handoff for request distribution (at very high load, stealing in HKS+WS rarely happens).
SHO achieves 10\% less peak throughput, because it is bottlenecked by the software handoff. In terms of 99th percentile, \ts{} does better than HKH at any load, with improvements reaching an order of magnitude as soon as the load exceeds 1 Mops.
HKH+WS and SHO start out with similar 99th percentile latencies as \ts{} under loads below 1 Mops, but under high load their 99th percentile latencies
rapidly deteriorate to reach values similar to HKH.
For an SLO on the 99th percentile latency of 50 $\mu$sec, i.e., 10 times the mean service time of a request,  \ts{} can perform 5.6 Mops, 2.4 times the throughput of its best competitor (HKH+WS). For an SLO of 100 $\mu$sec, \ts{} still achieves 1.75 times the throughput of its best competitor.

\ts{} achieves the best performance by overcoming the limitations of existing designs when dealing with variable-size items.
Early binding in HKH causes head-of-line blocking even at relatively low loads. At low or medium loads, work stealing mitigates head-of-line blocking in HKH, and brings $HKH+WS$ latencies close to that of late binding in SHO.  As the load increases, however, stealing occurs more rarely, and the performance of HKH+WS degrades to that of $HKH$. Late binding in SHO largely avoids head-of-line blocking, but sudden spikes of large requests hurt the high-percentile latencies of small requests. Further, the maximum throughput of SHO is bottlenecked by the maximum handoff rate sustainable by the handoff cores.

\pvs
~\\\noindent{\bf Latency of large requests.}
\ts{} leverages the insight that the latency of the slowest 1\% of the requests does not impact the 99th percentile. 
\ts{} restricts the 1\% largest requests to a subset of the cores, which may result in increased latencies for such requests.
We now evaluate  the performance penalty incurred by large requests in \ts{} as a consequence of size-aware sharding between small and large requests. Figure~\ref{fig:eval:large} reports the 99th percentile latency of large requests in \ts{} and HKH+WS (the best alternative).

Inevitably, \ts{} imposes some penalty on the performance of large requests under high load, reaching up to a factor of 2 for the 99th percentile latency of large requests before the system goes into saturation.
In this workload, large requests account to 0.125\% of the total, so the 99th percentile of large requests corresponds to 0.00125\% of the overall number of requests. We argue that moderately penalizing the very tail of the latency distribution is a reasonable price to pay for the order-of-magnitude improvement for the 99th percentile.

\begin{figure}[t!]  
\centering
       \includegraphics[scale = 0.55]{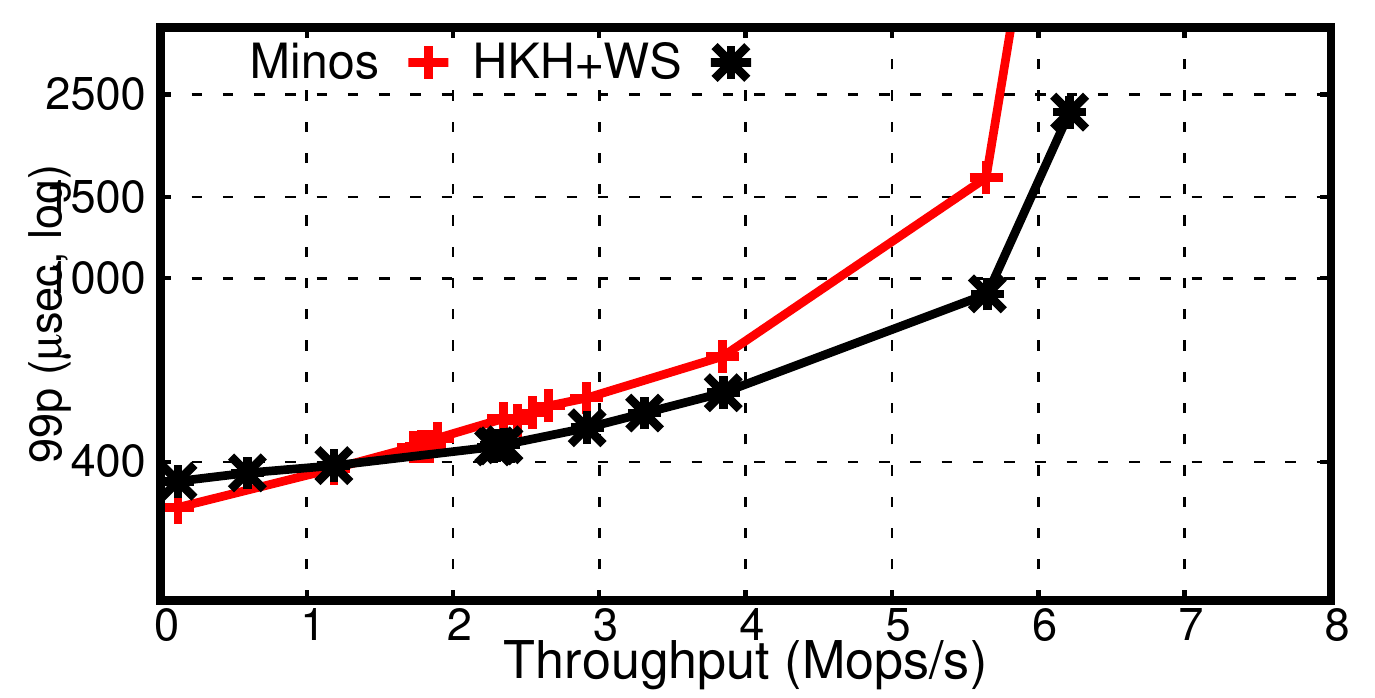}
       \caption{Throughput vs. 99th percentile latency of large requests with the default workload (y axis in log scale). \ts{} trades its large benefits in terms of the overall 99th percentile for a moderate penalty on a minority of large requests, which already represent a small fraction of the workload.}
        \label{fig:eval:large}
\end{figure}

\ts{} can improve the latency of large requests by allocating more cores to them. \ts{} currently determines the number of small cores by taking the ceiling of the total number of cores times the percentage of load generated by small requests. 
For this particular workload, it allocates only one core to the large requests.
This represents an over-allocation to small requests to completely isolate them from large requests, and hence an under-allocation for large requests. 
An alternative strategy is to allocate one more core to large requests, and let large cores steal from the RX queues of small ones to fully use any extra capacity. 
To avoid re-introducing head-of-line blocking, stealing can be done one request at a time, so that there is never a small request queued behind a large request.
We are currently experimenting with this alternative design, which would improve performance for large requests, while only introducing a small degradation for small requests.

\subsection{Write-intensive workload} 
\label{sec:eval:write}
We now investigate the effect of write intensity on \ts{}.
Figure~\ref{fig:eval:write} reports the 99th percentile of response times with all four systems and a 50:50 GET:PUT workload.

\ts{} continues to deliver a 99th percentile latency 1 order of magnitude lower than alternative approaches, up to the saturation point at 6.3 Mops, but overall achieves a lower (by 10\%) throughput than HKH and HKH+WS. 
Throughput values are in general higher than with the 95:5 workload, because replying to a PUT requires less network bandwidth, since the response message does not contain any item value payload.  This behavior is consistent with that observed by previous work~\cite{Lim:2014}. SHO is the only exception, as handoff cores represent the bottleneck.

A write-intensive workload shifts the bottleneck from the NIC to the CPU. \ts{} saturates the CPU earlier than HKH and HKH+WS because of the overhead stemming from profiling the workload and periodically aggregating them on core 0 to compute the 99th percentile of the item sizes. We are currently investigating techniques to reduce such overhead, e.g.,  sampling only a subset of the requests.
Alternatively, if traces of the target workload  are available for off-line analysis (as typical in production workloads~\cite{Atikoglu:2012,Nishtala:2013,Reda:2017}), the threshold between large and small requests can be set statically. With this variant, \ts{} is able to match the throughput of HKH and HKH+WS. 

\subsection{Sensitivity to item size distribution}
\label{sec:eval:variability}
We vary the percentage of large requests in the workload ($p_L$) and the maximum size of large requests ($s_L$). When changing the value of one, the other parameter keeps the default value. We then measure the maximum throughput achievable under different SLOs on the 99th percentile latency of 10 and 20 times the mean service time, i.e., $50 \mu sec$, and $100 \mu sec$. 

Figure~\ref{fig:eval:pl} and Figure~\ref{fig:eval:sl} report the increase in throughput achieved by \ts{} compared to the other designs (y axis in log scale).
Figure~\ref{fig:eval:pl} shows the results of the experiments in which we change $p_L$. Figure~\ref{fig:eval:sl} refers to changing $s_L$.
The graph on the left uses an SLO of 50 $\mu$sec, the one on the right 100 $\mu$sec.
When varying $p_L$, the maximum throughput achieved by \ts{} within the $50 \mu sec$  ($100\mu sec$) SLO ranges from 6.2 to 1.7 Mops (6.9 to 2.3 Mops), corresponding to $p_L = 0.0625$ and $p_L = 0.75$. 
When varying $s_L$, the maximum throughput achieved by \ts{} within the $50 \mu sec $  ($100\mu sec$) ranges from 6.2 to 4.7 Mops (6.9 to 4.7 Mops), corresponding to $s_L = 250 KB$ and $s_L = 1000KB$. 

\begin{figure}[t]
\centering
       \includegraphics[scale = 0.55]{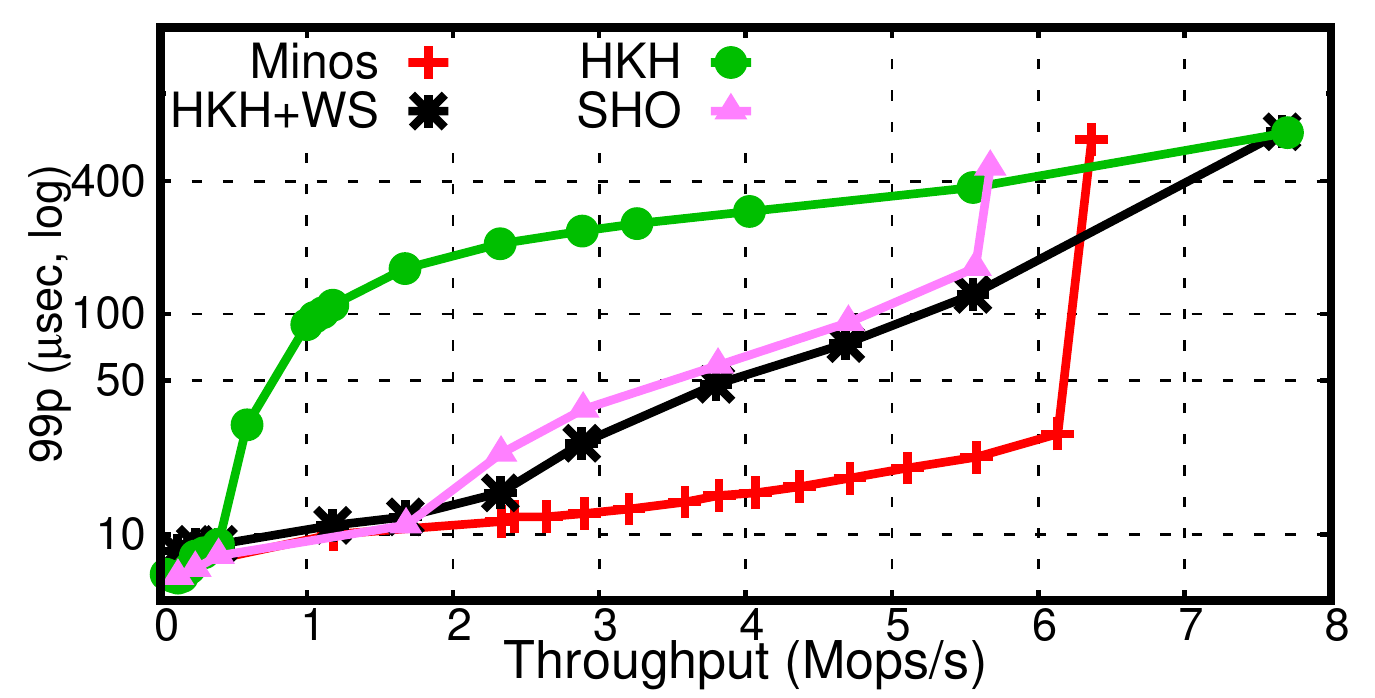}\caption{Troughput vs. 99th percentile latency for \ts{} vs. existing designs with the 50:50 GET:PUT workload (y axis in log scale).}
        \label{fig:eval:write}
\end{figure}

\begin{figure*}[t]
\begin{subfigure}
{0.48\textwidth}  
\centering
       \includegraphics[scale = 0.5]{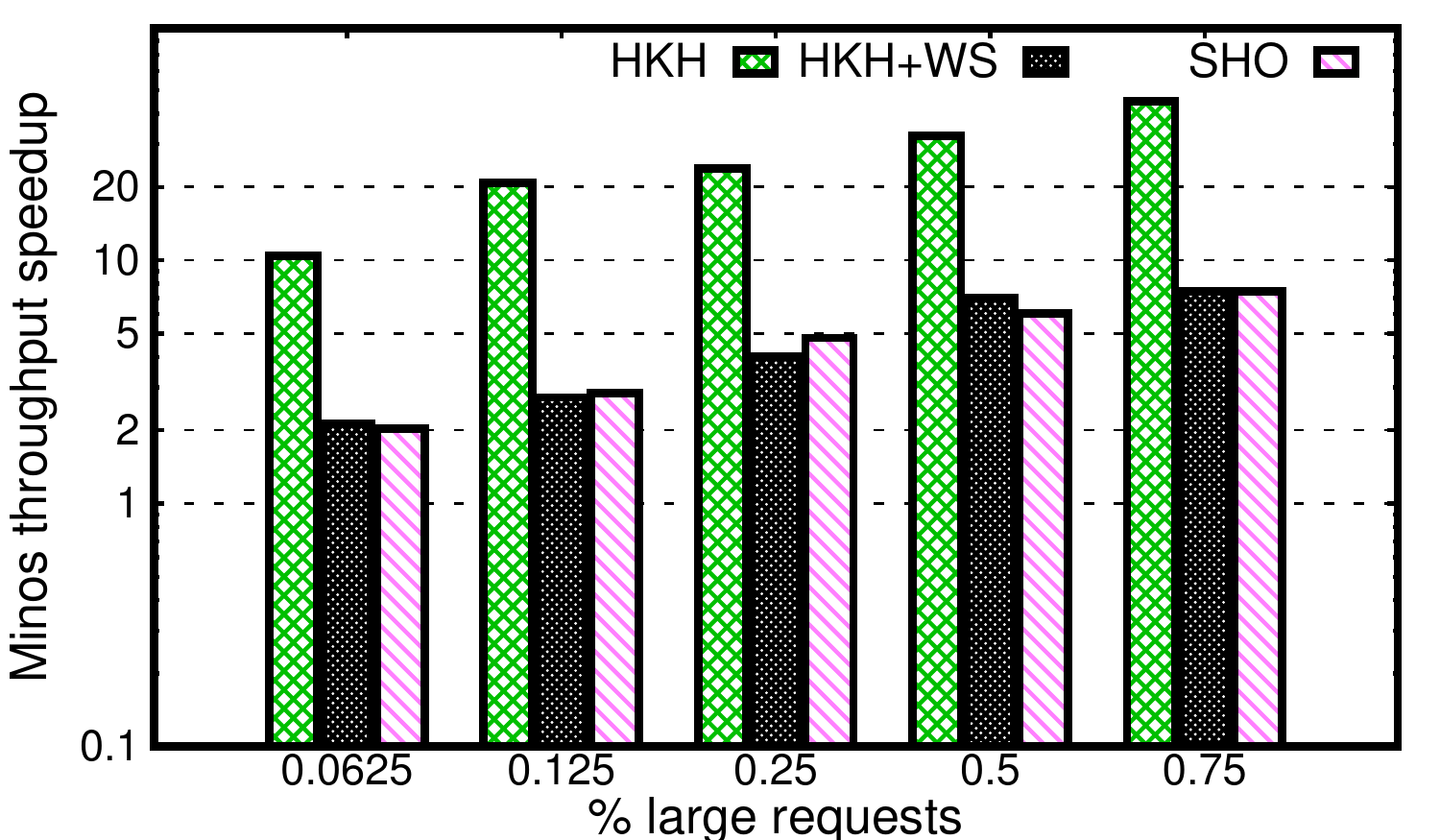}\caption{99p $\leq 50 \mu$sec.}\label{fig:eval:pl:50}
\end{subfigure}
\hfill
\begin{subfigure}{0.48\textwidth}
\centering
       \includegraphics[scale = 0.5]{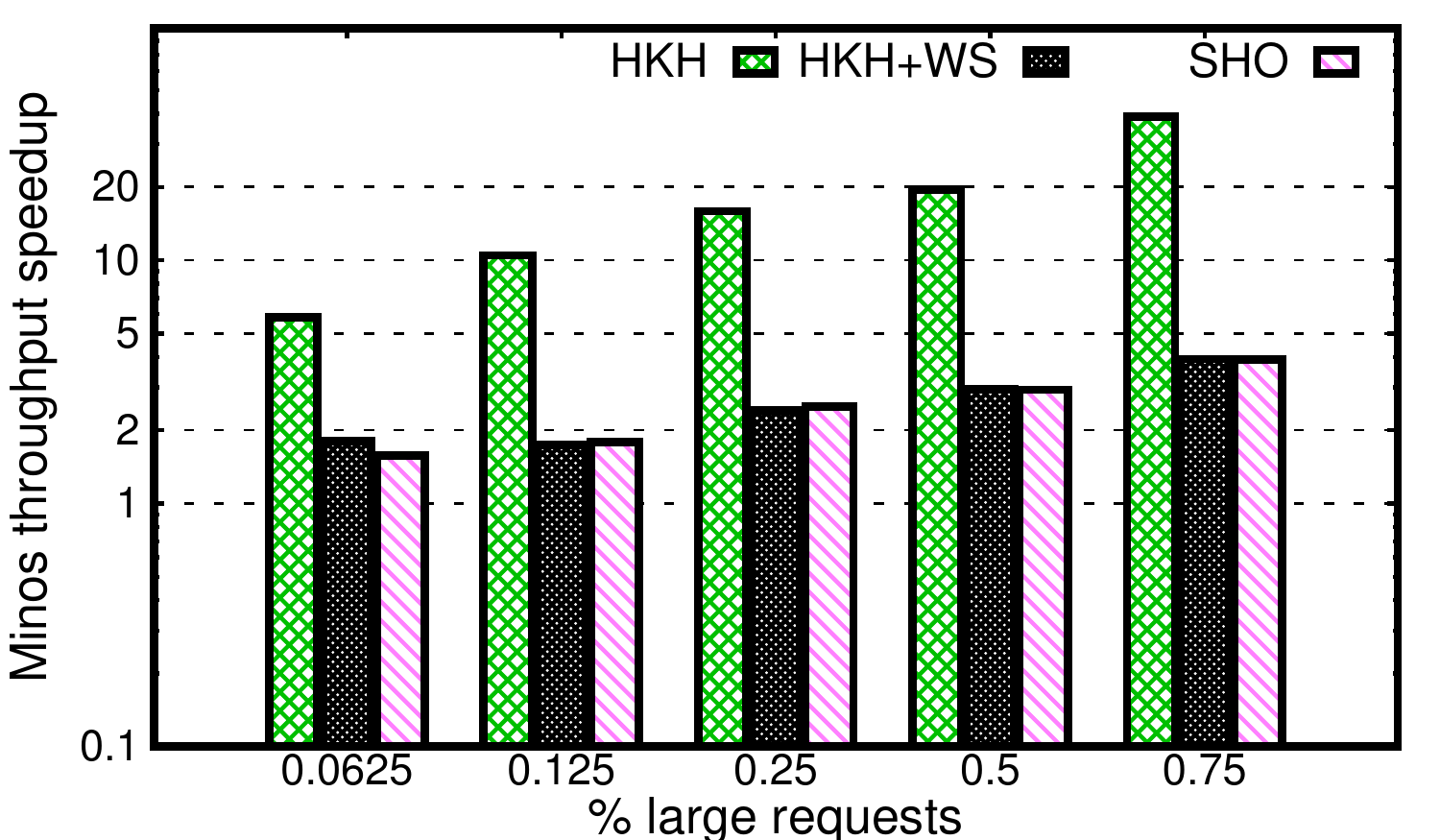}\caption{99p $\leq 100 \mu$sec.}\label{fig:eval:pl:100}
\end{subfigure}
\caption{Maximum throughput achievable for a given 99the percentile latency SLO with different percentages of large requests (y axis in log scale). Each bar represents the speedup of \ts{} over an alternative design (higher is better).}\label{fig:eval:pl}
\end{figure*}
\begin{figure*}[t]
\begin{subfigure}{0.48\textwidth} 
\centering
       \includegraphics[scale = 0.5]{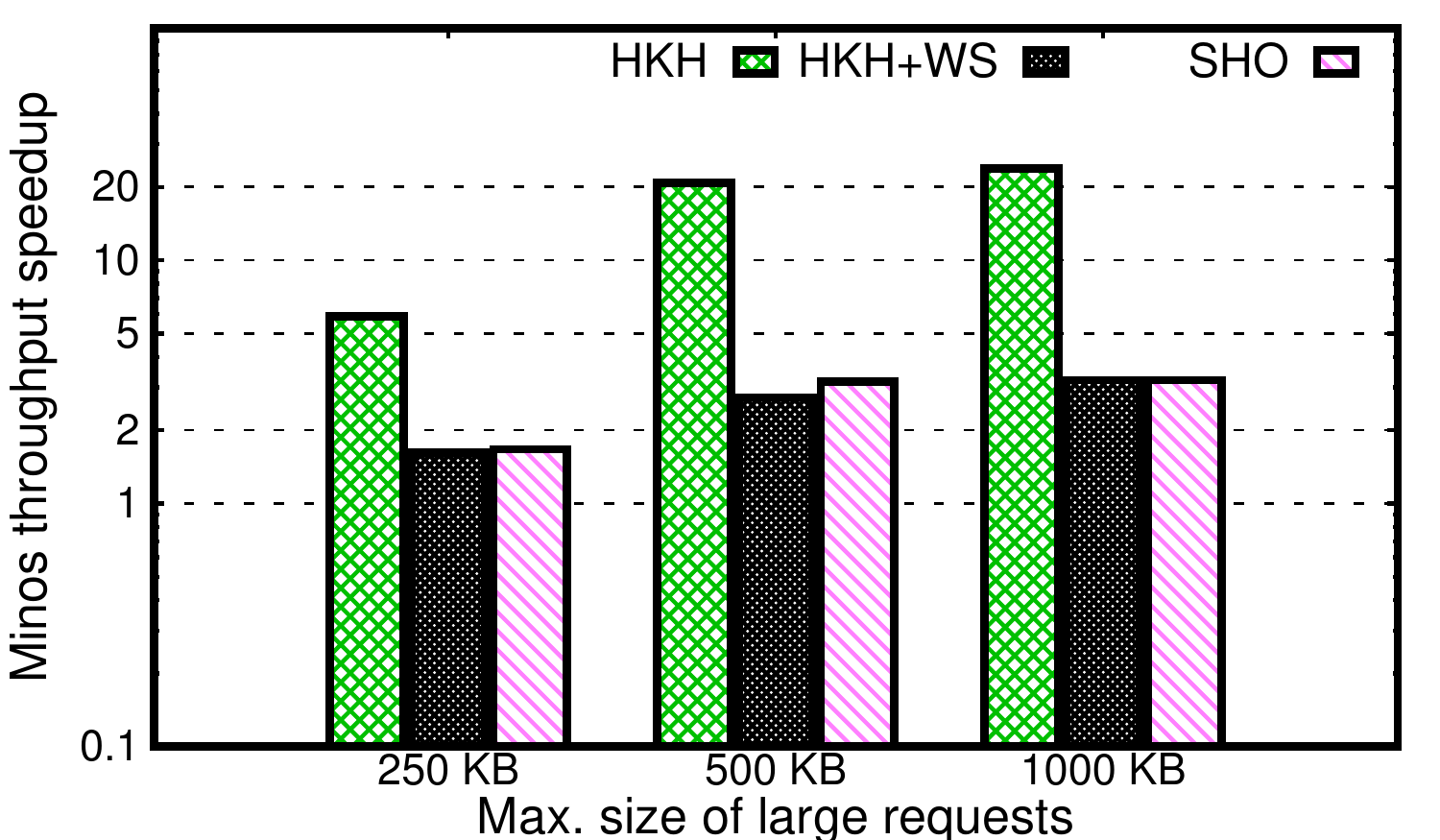}\caption{99p $\leq 50 \mu$sec.}\label{fig:eval:sl:50}
\end{subfigure}
\hfill
\begin{subfigure}{0.48\textwidth}
\centering
       \includegraphics[scale = 0.5]{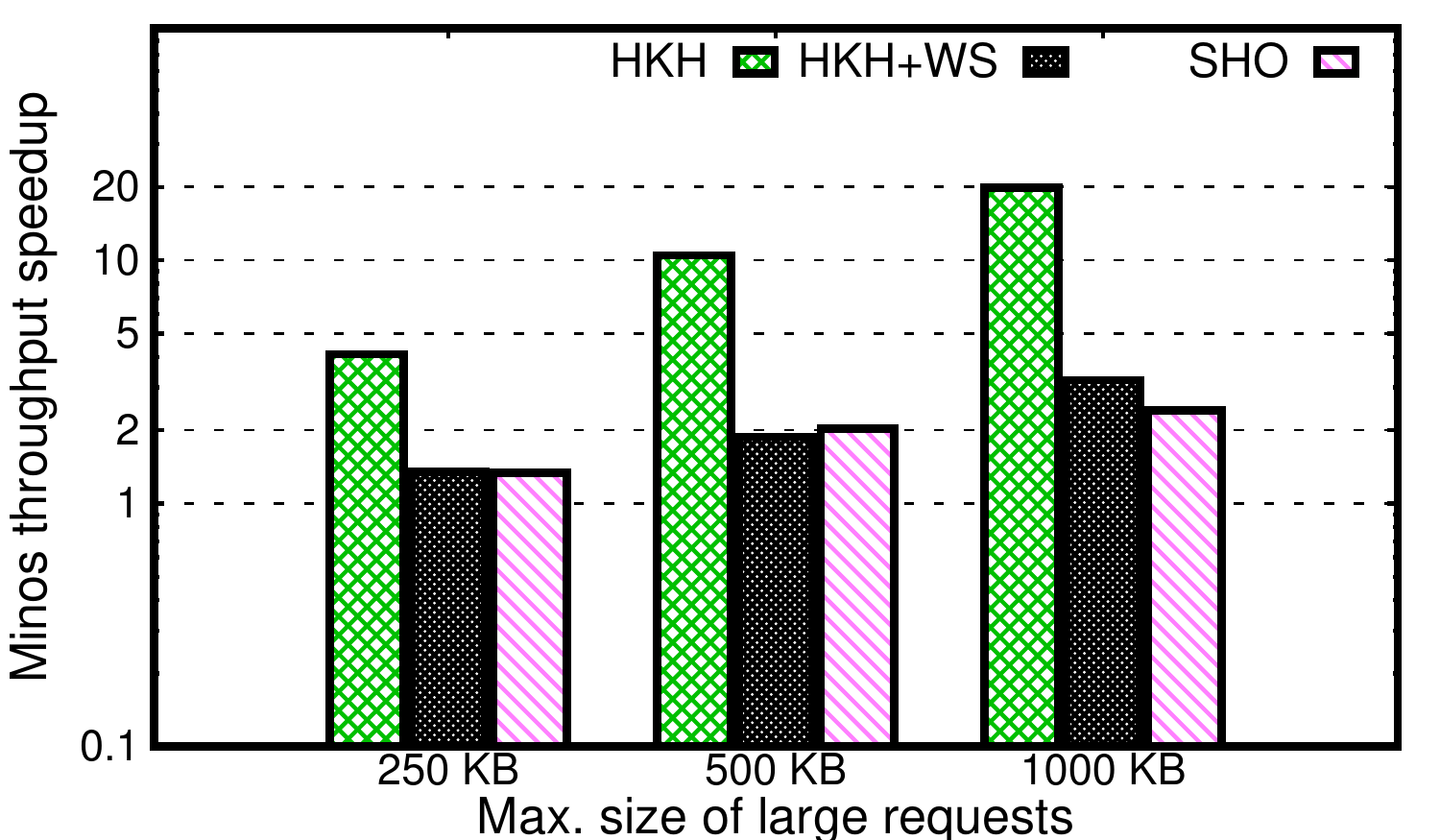}\caption{99p $\leq 100 \mu$sec.}\label{fig:eval:sl:100}
\end{subfigure}
        \caption{Maximum throughput achievable for a given 99th percentile latency SLO with different maximum sizes of large requests (y axis in log scale). Each bar represents the speedup of \ts{} over an alternative design (higher is better).} \label{fig:eval:sl}
\end{figure*}
\begin{figure*}[t]
\begin{subfigure}[b]{0.48\textwidth}  
\centering
       \includegraphics[scale = 0.55]{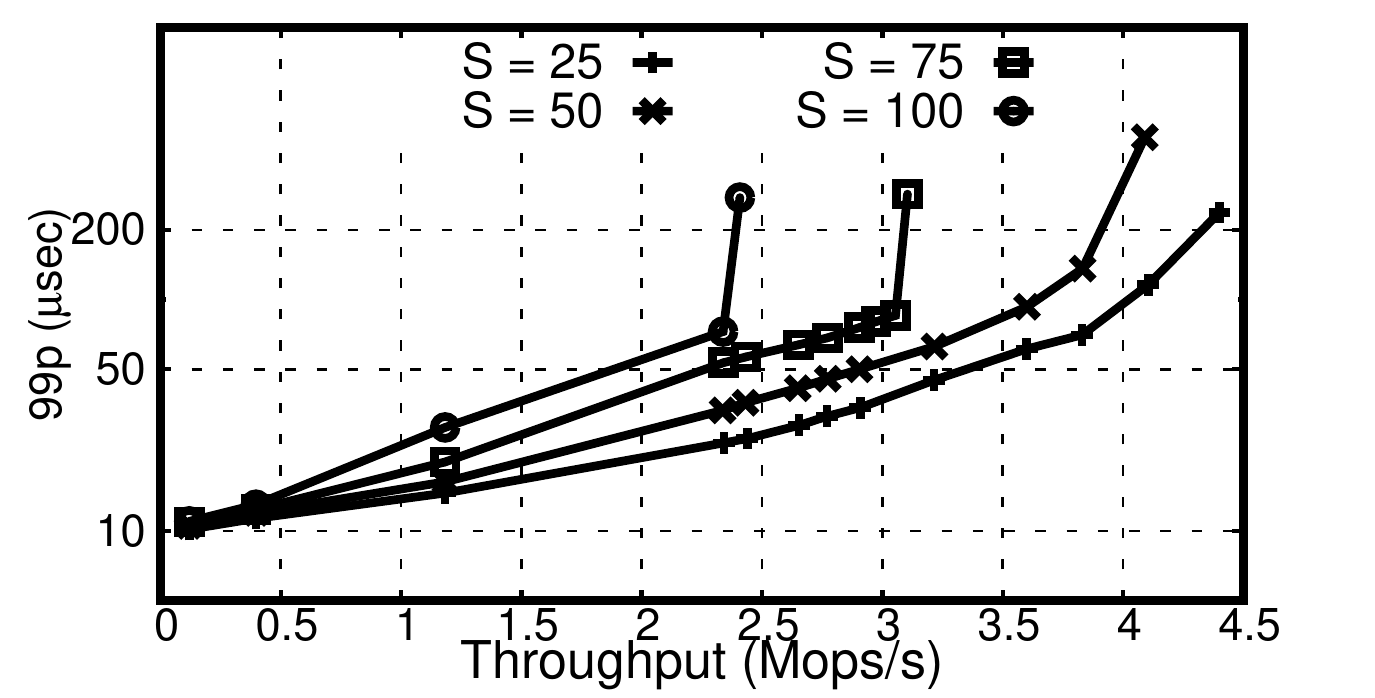}\caption{Throughput vs. 99th percentile latency}\label{fig:eval:sampling:resp}
\end{subfigure}
\hfill
\begin{subfigure}[b]{0.48\textwidth}
\centering
       \includegraphics[scale = 0.55]{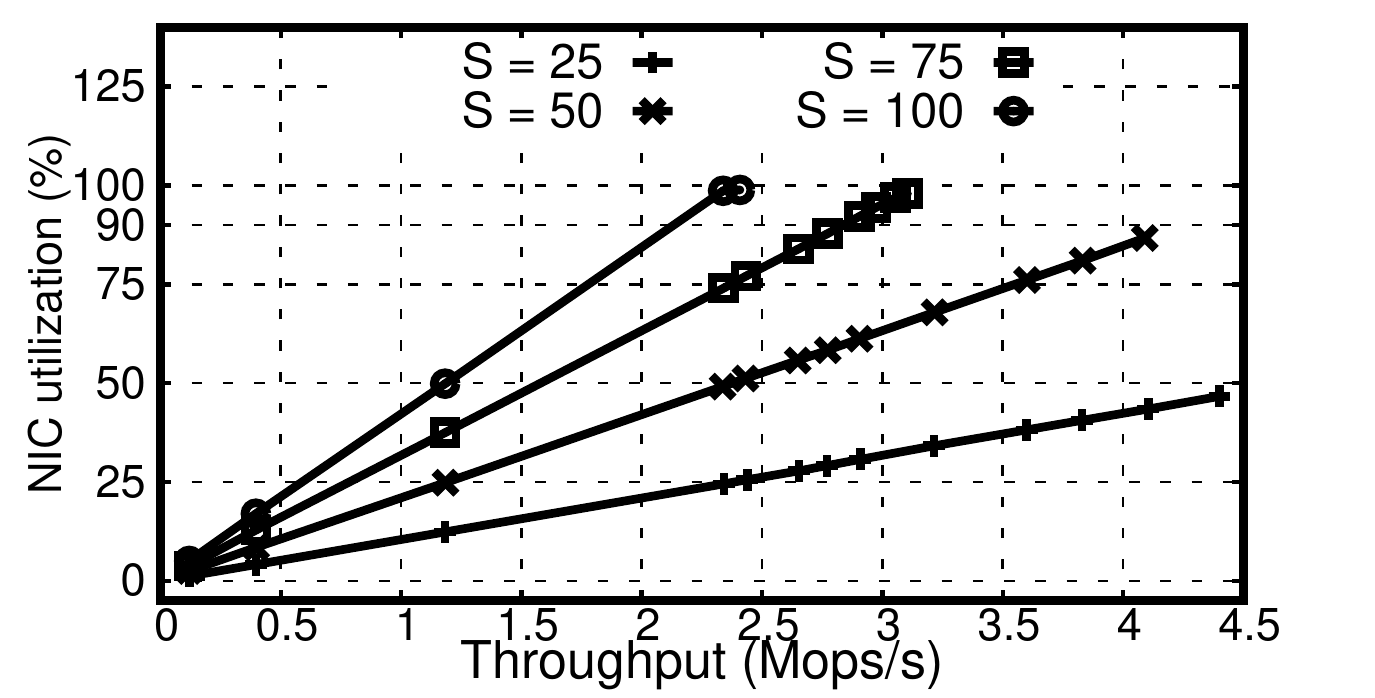}\caption{Throughput vs. NIC utilization.}\label{fig:eval:sampling:nic}
\end{subfigure}
        \caption{Scalability of \ts{} with more network bandwidth ($p_L = 0.75$). $S$ is the sampling percentage used to simulate more network bandwidth available. \ts{} processes and replies to a percentage $S$ of the requests. The remainder is processed, but the reply is dropped. \ts{} scales with more bandwidth (a) and saturates the NIC (b), except when the CPU becomes the bottleneck  ((b), S = 25).} \label{fig:eval:sampling}
\end{figure*}
\begin{figure*}[t!]
\begin{subfigure}[b]{0.48\textwidth} 
\centering
       \includegraphics[scale = 0.5]{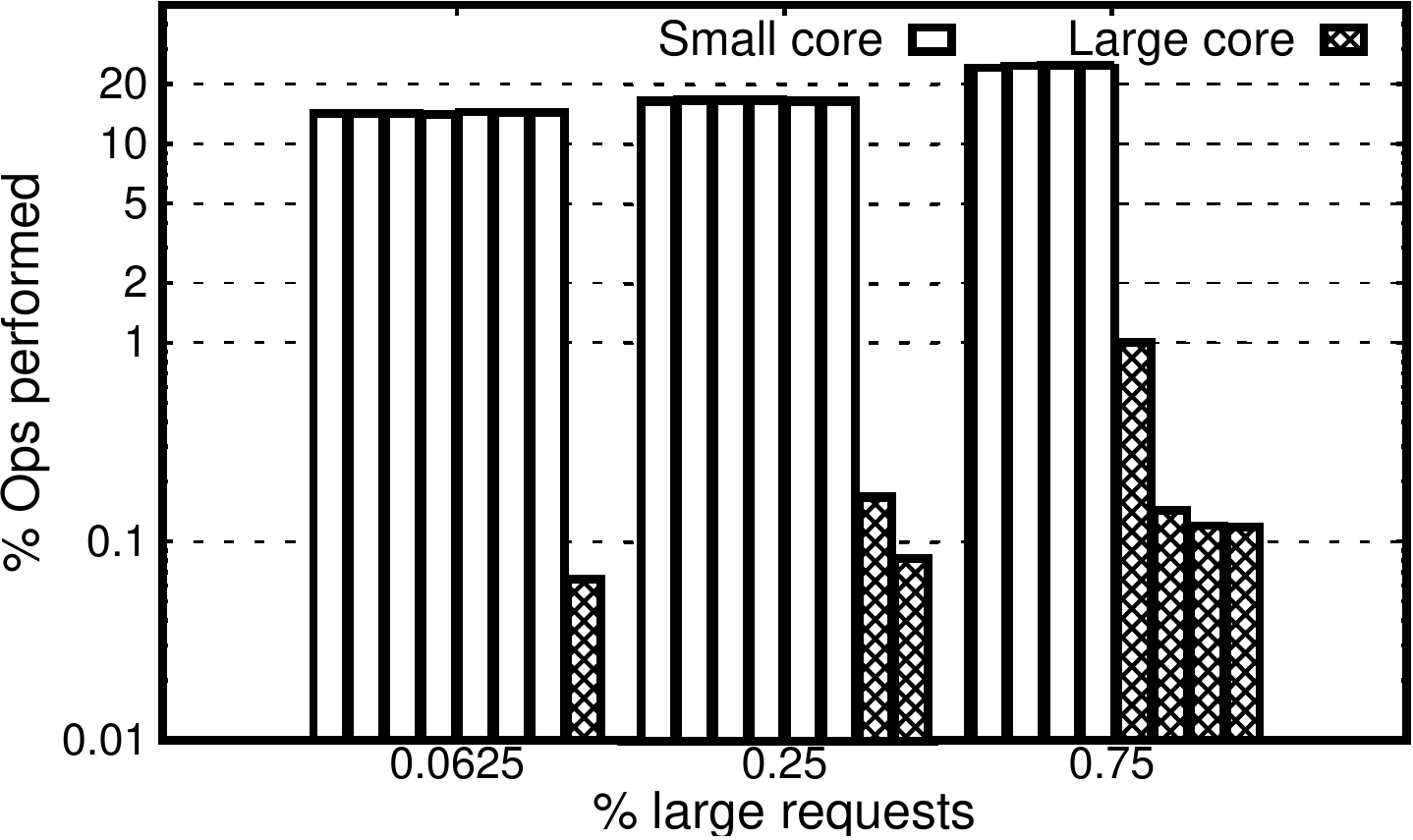}\caption{{\bf Operations} per second.}\label{fig:eval:lb:xput}
\end{subfigure}
\hfill
\begin{subfigure}[b]{0.48\textwidth}
\centering
       \includegraphics[scale = 0.5]{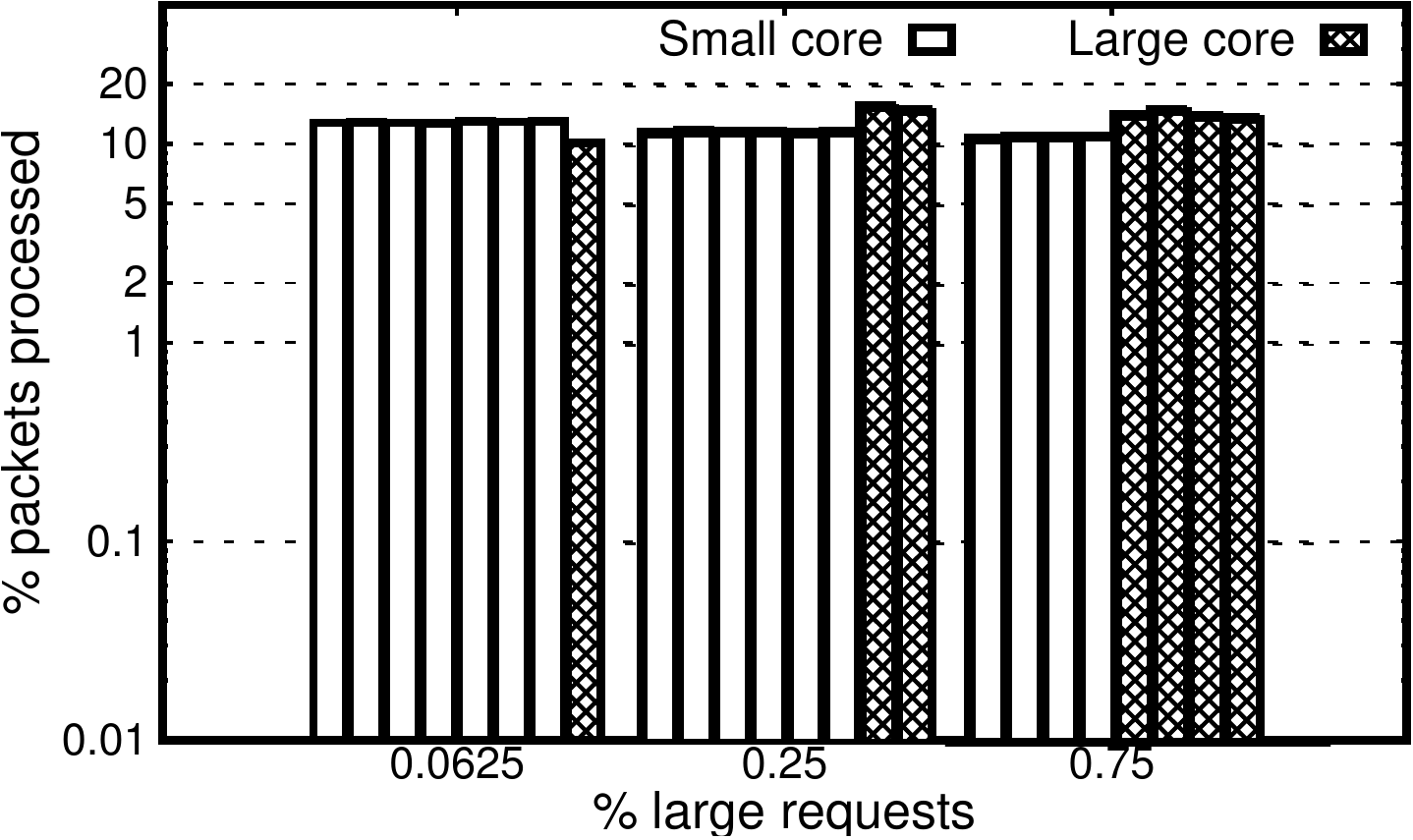}\caption{{\bf Packets} per second.}\label{fig:eval:lb:frames}
\end{subfigure}
\caption{Breakdown of the load per core in \ts{} (y axis in log scale). Large cores process fewer requests per second than small cores (a), but the number of packets processed per second is uniformly distributed across cores (b).}\label{fig:eval:lb}
\end{figure*}

\ts{} outperforms existing designs,  achieving consistently higher throughput for a given workload and a given SLO. The throughput speedup grows with $p_L$ and $s_L$, because 
the increased presence of large(r) requests negatively affects the latency of small requests, and hence the 99th percentile. 
As expected, the throughput gains are higher with the stricter SLO: the looser is the performance target, the smaller is the impact of \ts{}' design.
For the stricter SLO, \ts{} achieves a speedup of up to 7.4  w.r.t to HHK+WS (corresponding to the $p_L = 0.75$ case), i.e., the second best design. For the looser SLO, the speedup ranges from 1.34 ($s_L = 250$KB) to 3.9 ($p_L = 0.75$).

\subsection{Higher network bandwidth}
\label{sec:eval:bandwidth}
With the default workload, the NIC is 93\% utilized. 
With higher percentages of large requests, the system becomes network-bound.
In this section we investigate whether \ts{} can take advantage of larger network bandwidths.
Because we cannot provision our machines with more bandwidth, we shift the bottleneck from the NIC to the CPU by sampling the number of replies that the server sends back to clients.
That is, the server processes requests as before, up to the time at which it would otherwise send the reply to the client. Then, instead, it only sends replies to a percentage $S\%$ of the total requests, and drops the remaining ones. 
We vary $S$ from 100 to 25, and we measure the achieved performance (throughput and 99th percentile latency), as well as the utilization of the NIC. We choose the read-intensive workload with $p_L = 0.75$, as it quickly saturates the NIC when \ts{} replies to all requests.

Figure~\ref{fig:eval:sampling} reports the results of the experiment. The left plot shows the throughput vs. 99th percentile latency (y axis in log scale). The right one shows the utilization of the NIC as a function of the throughput. As $S$ decreases, \ts{} can sustain higher loads, because the bottleneck is increasingly shifted towards the CPU. \ts{} is able to fully utilize the available resources, by always reaching throughput values that bring either the NIC (S = 100,75,50) or the CPU (S = 25) close to saturation.

\subsection{Load balancing}
\label{sec:eval:lb}
\begin{figure}[b!]
\centering
       \includegraphics[scale = 0.55]{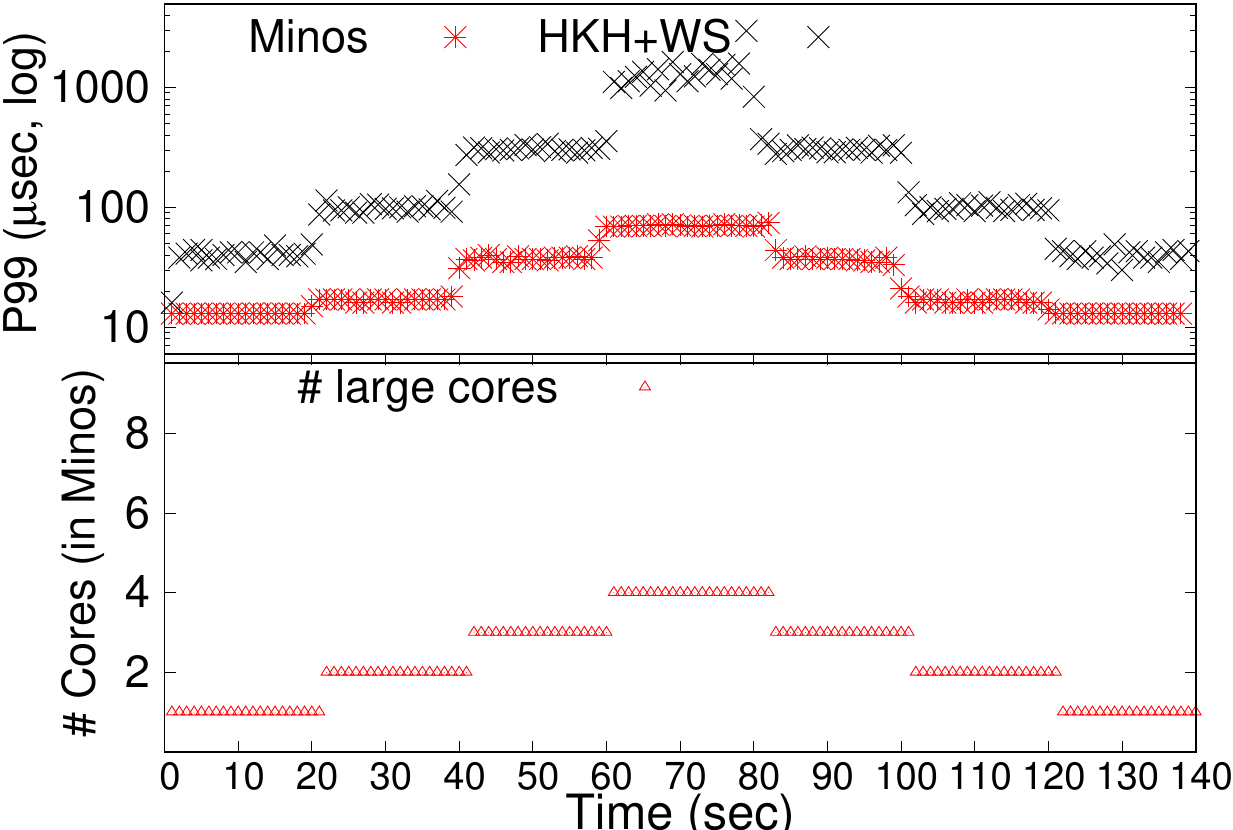}\caption{Evolution over time of the 99th percentile latency of \ts{} and HKH+WS with a dynamic workload (top, with y axis in log scale) and evolution over time of number of large cores in \ts{} (bottom). Every 20 seconds the percentage of large requests changes, first growing from 0.125 to 0.75 and then shrinking back. \ts{} adapts to changing workload conditions and delivers up to two orders of magnitude lower 99th percentile latencies.}\label{fig:eval:dyn}
\end{figure}

We now evaluate the ability of \ts{} to distribute the load evenly across cores, for a variety of workloads. 
To this end, we measure the load sustained by each core with $p_l$ = 0.0625, 0.25, 0.75, corresponding to low, medium and high load posed by large requests.
Figure~\ref{fig:eval:lb:xput} reports the percentage of requests performed, and Figure~\ref{fig:eval:lb:frames} reports the percentage of packets processed by each core (y axis in log scale).
Two conclusions can be drawn.
First, all cores process roughly the same number of packets, and hence roughly perform the same amount of work.
Small cores obviously process more requests per second, as these requests involve less work. Large cores process different requests per second among each other, as a consequence of the size-aware sharding that \ts{} implements also within large requests.
Second, \ts{} varies the number of small and large cores as a function of the workload, such that the work is balanced among all cores.

\subsection{Dynamic workload}
\label{sec:eval:dynamic}
We finally demonstrate the capability of \ts{} to adapt to changing workloads. To this end, we run a workload in which the percentage of large operations $p_L$ varies every 20 seconds. It first grows gradually from 0.125 to 0.75, and then shrinks back to 0.125. We keep the request arrival rate fixed at 2.25 Mops, corresponding to high load for $p_L = 0.75$.
Figure~\ref{fig:eval:dyn}(top) compares the performance achieved by \ts{} and HKH+WS, i.e., the second best design. Each point represents the 99th percentile latency as measured over a 1 second window (y axis in log scale). 
Figure~\ref{fig:eval:dyn}(bottom) shows how many cores \ts{} assigns to large requests over time.
\ts{} achieves latencies up to 2 orders of magnitude lower than HKH+WS ($\approx 70 \mu$sec vs $\approx 1 msec$ with $p_L = 0.75$). \ts{} achieves this result by programmatically  allocating cores to small and large requests proportionally to their corresponding loads. 

\section{Related Work}
\label{sec:rw}
To the best of our knowledge, \ts{} is the first KV store to introduce the concept of size-aware sharding to address the challenges of delivering $\mu$sec-scale tail latency in presence of item size variability. We now discuss related systems.

\pvs
~\\\noindent{\bf In-memory KV stores.} A plethora of in-memory KV stores have been proposed in the last years. These systems propose different designs based on new data-structures (CPHash~\cite{Metreveli:2012}, Masstree~\cite{Mao:2012}, MemC3~\cite{Fan:2013}) and lightweight network stacks (Chronos~\cite{Kapoor:2012}, MICA~\cite{Lim:2014,Li:2015}, RamCloud~\cite{Ousterhout:2015}, RockSteady~\cite{Kulkarni:2017}), or on the use of RDMA (Pilaf~\cite{Mitchell:2013}, Herd~\cite{Kalia:2014}, FaRM~\cite{ Dragojevic:2014}, RFF~\cite{Su:2017}, FaSST~\cite{Kalia:2016}), FPGAs (KV-Direct~\cite{Li:2017}),  GPUs (Mega-KV~\cite{Zhang:2015}, MemcacheGPU~\cite{Hetherington:2015}), HTMs (DrTM~\cite{Wei:2015,Chen:2016}), or other specialized hardware (~\cite{Kaufmann:2016,Blott:2015}).

None of these systems addresses the problem of achieving low tail latency in presence of item size variability, which is the primary focus of \ts{}. In addition, \ts{} only assumes the availability of commodity hardware. Investigating the synergies between the design of \ts{} and specialized hardware is an interesting avenue for future work.

\pvs
~\\\noindent{\bf Size-aware data-stores.} We are aware of a few data stores that take into account the size of items or requests to improve performance. Rein~\cite{Reda:2017} supports multi-key get requests and processes them taking into account the number of keys involved in a request. Rein relies on the assumption that there is only a weak correlation between the size of an item and the service time of a request for that item. \ts{}, instead, targets workloads with high item size variability, for which the service time of a request strongly depends on the size of the corresponding item (see Figure~\ref{fig:background:service}). 

AdaptSize~\cite{Berger:2017} is a caching system for content delivery networks that reduces the probability of caching large objects, so as to increase the hit rate of smaller, more frequently accessed ones. AdaptSize targets a problem that is orthogonal to \ts{}, which assumes the presence in memory of both small and large items.

Other data stores for non-homogeneous requests~\cite{Harchol-Balter:2003,Zhang:2005,Ciardo:2001} target static content and leverage the {\em a priori} presence of a central component (the Linux kernel on a single-core architecture~\cite{Harchol-Balter:2003} or a scheduler in a distributed system~\cite{Ciardo:2001,Zhang:2005}) to implement request scheduling. By contrast, \ts{} deals with mixed read/write workloads and is suited for multi-core architectures with multi-queue NICs.

\pvs 
~\\\noindent{\bf Operating systems.} IX~\cite{Belay:2016} and ZygOS~\cite{Prekas:2017} use lightweight network stacks to support applications with $\mu$sec-scale SLOs. ZygOS implements work stealing to avoid core idleness and reduce head-of-line blocking. As we show by means of simulation (\S~\ref{sec:background:hob}) and experimental data (\S~\ref{sec:eval}), this approach cannot fully avoid head-of-line blocking as done by \ts{}, because work stealing $i)$ is agnostic of the CPU time corresponding to serving a request; and $ii)$ is only triggered by idle cores, whose presence becomes less likely as the load increases.

\pvs
~\\\noindent{\bf Job schedulers.} There is a vast literature on scheduling techniques for cluster and data center jobs of heterogeneous size~\cite{Harchol-Balter:2013:book}. Proposed approaches include workload partitioning~\cite{Crovella:1998,Harchol-Balter:2003,Delgado:2016}, pre-empting~\cite{Bansal:2001} or migrating large jobs~\cite{Harchol-Balter:2002,Haque:2017}, and stealing~\cite{Delgado:2015,Li:2016}. Similar techniques have been applied also in the context of network flow scheduling~\cite{Guo:2001,Hong:2012}.

\ts{} draws from these techniques  to efficiently implements size-aware request sharding in an in-memory key-value store, so as to avoid head-of-line blocking and achieve load balancing.

\section{Conclusion}
\label{sec:conclusion}
This paper presents \ts{}, an in-memory key-value store designed to deliver $\mu$sec-scale tail latency with workloads characterized by highly variable item sizes, as frequent in production workloads.
\ts{} implements size-aware sharding, a new technique that assigns small and large requests to disjoint set of cores. This ensures small requests never wait due to the collocation with a long request. \ts{} identifies at runtime the size threshold between long and short requests, and the amount of cores to allocate to them, so as to achieve low 99th percentile latency. 
We compare \ts{} to three state-of-the-art designs and we show that, compared to its closest competitor, \ts{} achieves a 99th percentile latency that is up to two orders of magnitude lower. Put differently, for a given value for the 99th percentile latency equal to 10 times the mean service time, \ts{} achieves a throughput that is up to 7.4 times higher.

\bibliographystyle{ACM-Reference-Format}

\end{document}